\documentclass[sigconf]{acmart}
\usepackage{subcaption}
\usepackage{graphicx}
\usepackage{lipsum}
\usepackage{breqn}
\usepackage{tabularx}
\usepackage{array}
\usepackage{colortbl}
\usepackage[utf8]{inputenc}
\def\BibTeX{{\rm B\kern-.05em{\sc i\kern-.025em b}\kern-.08emT\kern-.1667em\lower.7ex\hbox{E}\kern-.125emX}}

\copyrightyear{2020}
\acmYear{2020}
\setcopyright{acmlicensed}
\acmBooktitle{}
\acmPrice{}
\acmDOI{}
\acmISBN{}
\newcommand{\etal}{\textit{et al.}}
\begin{document}

%
% The "title" command has an optional parameter, allowing the author to define a "short title" to be used in page headers.
\title[Towards a ML-Driven Trust Evaluation Model for SIoT: A Time-aware Approach]{Towards a Machine Learning-driven Trust Evaluation Model for Social Internet of Things: A Time-aware Approach}

\author{Subhash Sagar, Adnan Mahmood, Quan Z. Sheng, and Munazza Zaib}
\affiliation{%
  \institution{Department of Computing, Macquarie University}
  \city{Sydney, NSW 2109}
  \country{Australia}}

\author{Wei Emma Zhang}
\affiliation{%
  \institution{School of Computer Science, The University of Adelaide}
  \city{Adelaide, SA 5005}
  \country{Australia}}

% \author{Subhash Sagar$^1$, Adnan Mahmood$^1$, Quan Z. Sheng$^1$, Munazza Zaib$^1$, and Wei Emma Zhang$^2$}
% \affiliation{\vspace{0.1cm}
% \institution{$^1$ Department of Computing, Macquarie University, Sydney, NSW 2109, Australia}
% \institution{$^2$ School of Computer Science, The University of Adelaide, Adelaide, SA 5005, Australia}
% }
% \email{{subhash.sagar, adnan.mahmood, munazza-zaib.ghori}@hdr.mq.edu.au}
% \email{wei.e.zhang@adelaide.edu.au, michael.sheng@mq.edu.au}
% \author{Subhash Sagar\affmark[1], Adnan Mahmood\affmark[1], Munazza Zaib, Quan Z. Sheng\affmark[1], and Wei Emma Zhang\affmark[2]}
%  \affiliation{%
%   \affaddr{\affmark[1]Department of Computing, Macquarie University, Sydney, NSW 2109, Australia}
%  }
%   \affiliation{%
%   \affaddr {\affmark[2]School of Computer Science, The University of Adelaide, Adelaide, SA 5005, Australia}
%  }
%  \email{{subhash.sagar, adnan.mahmood, munazza-zaib.ghori}@hdr.mq.edu.au}
%  \email{wei.e.zhang@adelaide.edu.au; michael.sheng@mq.edu.au}

\vspace{2em}

%
% By default, the full list of authors will be used in the page headers. Often, this list is too long, and will overlap
% other information printed in the page headers. This command allows the author to define a more concise list
% of authors' names for this purpose.
%\renewcommand{\shortauthors}{Zawar et al.}

%
% The abstract is a short summary of the work to be presented in the article.
\begin{abstract}
The emerging paradigm of the Social Internet of Things (SIoT) has transformed the traditional notion of the Internet of Things (IoT) into a social network of billions of interconnected smart objects by integrating social networking facets into the same. In SIoT, objects can establish social relationships in an autonomous manner and interact with the other objects in the network based on their social behaviour. A fundamental problem that needs attention is establishing of these relationships in a reliable and trusted way, i.e., establishing trustworthy relationships and building trust amongst objects. In addition, it is also indispensable  to ascertain and predict an object's behaviour in the SIoT network over a period of time. Accordingly, in this paper, we have proposed an efficient time-aware machine learning-driven trust evaluation model to address this particular issue. The envisaged model deliberates social relationships in terms of friendship and community-interest, and further takes into consideration the working relationships and cooperativeness (object-object interactions) as trust parameters to quantify the trustworthiness of an object. Subsequently, in contrast to the traditional weighted sum heuristics, a machine learning-driven aggregation scheme is delineated to synthesize these trust parameters to ascertain a single trust score. The experimental results demonstrate that the proposed model can efficiently segregates the trustworthy and untrustworthy objects within a network, and further provides the insight on how the trust of an object varies with time along with depicting the effect of each trust parameter on a trust score.  
\end{abstract}


\begin{CCSXML}
<ccs2012>
  <concept>
      <concept_id>10002978.10003014.10003017</concept_id>
      <concept_desc>Security and privacy~Mobile and wireless security</concept_desc>
      <concept_significance>300</concept_significance>
      </concept>
  <concept>
      <concept_id>10002978.10002997</concept_id>
      <concept_desc>Security and privacy~Intrusion/anomaly detection and malware mitigation</concept_desc>
      <concept_significance>300</concept_significance>
      </concept>
 </ccs2012>
\end{CCSXML}

\ccsdesc[300]{Security, Privacy, and Trust~Trust and Network Security}
\ccsdesc[300]{Security and Privacy~Intrusion/Anomaly Detection and Malware Mitigation}

%
% Keywords. The author(s) should pick words that accurately describe the work being
% presented. Separate the keywords with commas.
\keywords{Machine Learning, Social Internet of Things, Trustworthiness Management, Social Similarity, Cooperativeness, Friendship, Community-of-Interest
}

%
% A "teaser" image appears between the author and affiliation information and the body 
% of the document, and typically spans the page. 

%
% This command processes the author and affiliation and title information and builds
% the first part of the formatted document.
\maketitle

\section{Introduction}

Rapid advancements in communication and computing technologies have let to the evolution of billions of smart objects (e.g., smart machines, smartwatch, and smart cars) equipped with sensing, processing, communication capabilities that not only enables an object-object communication across the internet but also forms the well-known paradigm of the Internet of Things (IoT) \cite{fortino2014} \cite{ATZORI20102787}. This paradigm has undeniably produced enormous business, and opened doors to many applications and services in various valuable sectors \cite{Khet2016}, and led to a tremendous increase in connected smart objects that are expected to surpass 75.44 billion by 2025 \cite{iot-devices}. Furthermore, over the past decade, numerous research endeavors have analyzed the possibilities of integrating the notion of social networking into the IoT ecosystem. This integration has led to a new paradigm of the Social Internet of Things (SIoT), wherein, objects and humans are maintained at social and physical level (the left part of Figure 1) in order to facilitate the owners of the objects to set the rules for protecting their privacy. Moreover, each object in SIoT is capable of establishing relationships (as illustrated in Figure 2) with other objects in terms of their ownership (relation with the objects belonging to the same owner), social relationship (relation with the objects belonging to the friends in social network), parental relationship (relation with the objects from the same manufacturer), co-location and co-work relationships (relation with the objects from the same location and work environment), and enables autonomous inter-object interaction based on these relationships \cite{ATZORI20123594}\cite{MS201932}.

\begin{figure*}[t]
\centering
\includegraphics[width=0.8\linewidth]{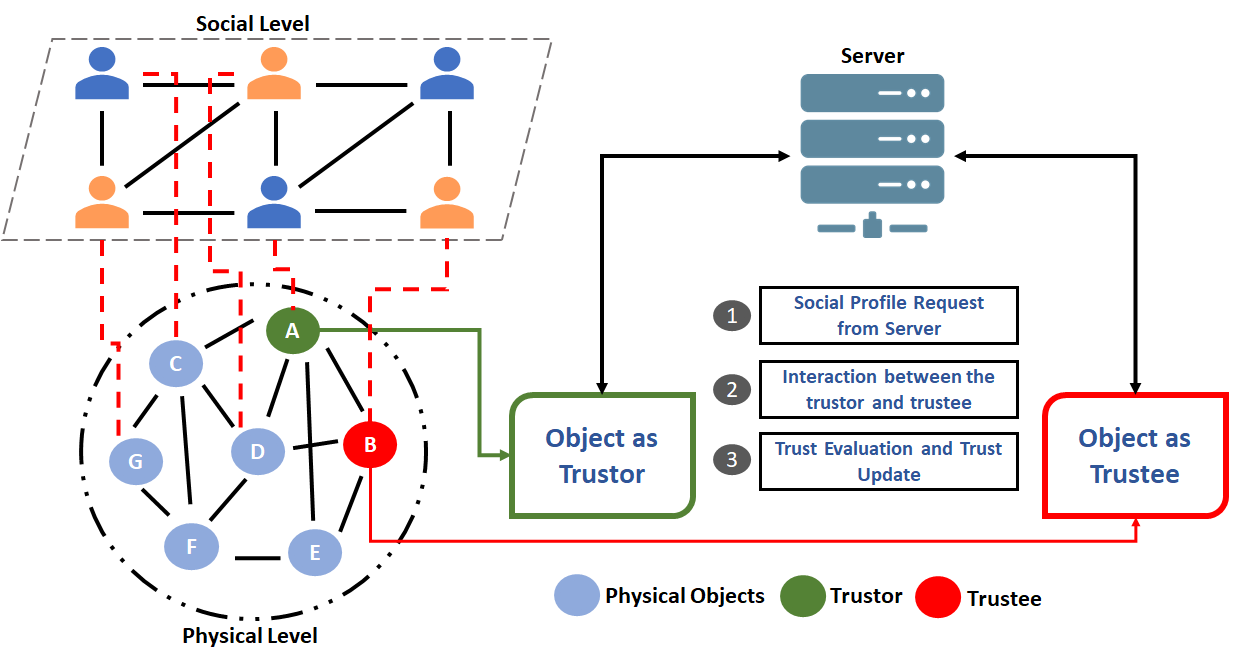}
\caption{Illustration of SIoT components and process of trust evaluation} \label{fig:trust_model}
\end{figure*}

Trust plays an important role in establishing and maintaining social relationships since social objects only involve themselves in a relationship when the participating objects are trustworthy enough to reduce the probable risk in decision-making \cite{ATZORI20123594}. Trustworthy relationships among the objects make it easy for them to only respond to the service requests from the familiar objects in the network, thereby, reducing the exposure to malicious objects \cite{AN2013799}.

The notion of trust has been studied in many disciplines, i.e., sociology, psychology, and computer science \cite{6362662}\cite{sociology}, nevertheless, the perception of trust in each of these disciplines is different, and therefore, there is no generally acknowledged definition of trust. As a result, it is essential to look at trust from the SIoT point of view. In SIoT, trust is characterized as the ``confidence'' of a trustor in a trustee to achieve an objective under a particular setting inside a particular timespan. Trust is a dynamic process that involves trustor, trustee, and the underlying context. The trust computation process in this paper follows three steps: 1) social relationships request, 2) interactions in terms of the number of successful packets transmission, and 3) trust evaluation by quantifying the trust metrics from step 1 and step 2 and trust update. The right part of Figure 1 illustrates the trust process by utilizing the services of a local authority as a server, wherein, if an object $A$ needs to compute the trustworthiness of an object $B$, it will first request the social profile (i.e. friends, interest groups, and working relationship) of the object $B$ from the server. Subsequently, after the interaction, object $A$ evaluate and update the trustworthiness of object $B$ in the network through a local authority. 

\begin{figure}[h!]
\centering
\includegraphics[width=\linewidth]{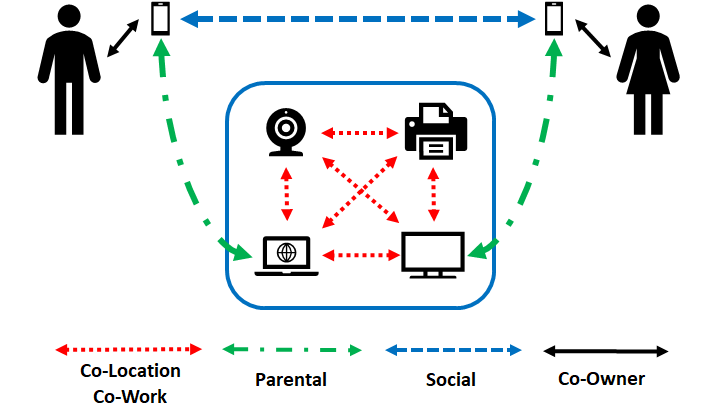}
\caption{SIoT relationships} \label{fig:trust_model}
\end{figure}

As of recently, research into SIoT primarily focused on the definition and construction of these SIoT relationships, nevertheless, the SIoT paradigm still lacks in some fundamental aspects such as understanding how these relationships can be used and quantified to build a reliable trustworthy system based on the behaviour of the objects \cite{8737491}. Based on these observations, in this work, we hereby propose a trust evaluation model, wherein, the trustworthiness of an object is computed dynamically by employing social relationships as well as the interactions among the objects. Overall, the main contributions of this paper are summarized as follows:

\begin{itemize}
\item[1)] A time-aware trust evaluation model has been proposed that ascertains the trust of an object by exploiting social relationships (i.e., friendship and community-of-interest), co-work relationship, and cooperativeness in terms of interactions among the objects. Moreover, both the direct trust and reputation (indirect trust) are employed to identify between a trustworthy and, an untrustworthy node. 
\item[2)] In contrast to the traditional aggregation techniques, a machine learning-driven aggregation scheme has been envisaged to aggregate all the trust metric for a single trust score; and,
\item[3)] Finally, the experimental evaluation of the proposed model gives insight on how the trustworthiness of an object varies with respect to time along with depicting the effect of each trust metric while computing the trustworthiness of an object.  
\end{itemize}

The remainder of this paper is organized as follows. The recent work in the literature is discussed in Section 2. Section 3 formally defines the problem and introduces the proposed trust evaluation model. Section 4 provides the detailed simulation setup and the machine learning-driven aggregation scheme. Results and discussion are described in Section 5. Finally, Section 6 concludes the paper and discusses the future work.

\section{State-of-the-Art}

Recently, there has been an increasing interest by a number of researchers to model the trustworthiness management in SIoT since it is one of the fundamental concerns in SIoT research \cite{7097037}\cite{8653859}\cite{6940301}\cite{sagarGlobecom}. A wide range of facets can be contemplated for ascertaining the trustworthiness, including but not limited to, social relationships, reputation, quality-of-service, context (i.e., energy and time), etc. In \cite{7097037},  Boa {\etal} proposed a trust evaluation mechanism, wherein, three trust parameters are considered to evaluate the trustworthiness of an object. The considered parameters are cooperativeness, honesty, and community-of-interest, and to aggregate these parameters, an adaptive weighted sum technique has been suggested. Moreover, their trust model considers direct trust and indirect trust (as recommendations) while quantifying the trust of an object.

Anuoluwapo {\etal} \cite{8653859} suggested a dynamic trust evaluation model CTRUST, wherein, objects decide the functional parameters in terms of collaborative context, to estimate the trustworthiness of an object in the SIoT network. Besides, CTRUST includes the decay factor to ensure the trust degradation process more efficiently and in a consistent manner. Nevertheless, weighted trust parameters have been equipped to aggregate the final trust. Each node decides the weights of each parameter at run-time to adjust the trustworthiness of an object in various contexts. Moreover, an adaptive trust protocol for service-oriented architecture is proposed in \cite{6940301}, wherein social similarity and trust feedback are considered to measure the trustworthiness of an object, and the protocol's effectiveness is demonstrated through service-oriented applications. However, an object in the discussed models need to discover the best trust parameters and their respective weights as per the environmental conditions.    

There are numerous research works on the trust computational model delineated in the literature targeting different applications of IoT such as vehicular networks \cite{8887207}\cite{8730675}, peer-to-peer networks \cite{1318566}\cite{socialtrust}, recommendation systems \cite{8936977}\cite{7845500}, and mobile crowdsourcing \cite{8662641}\cite{7195560}. Alnasser {\etal} \cite{8887207} proposed a recommendation-based trust model for vehicle-to-everything (V2X) communication where direct trust and recommendations from credible vehicles are combined to obtain the trust score of a vehicle. Subsequently, a dynamic adaptive weighted sum method is used to aggregate both direct trust and recommendations. Yuanyi {\etal} \cite{8936977} presented a time-aware smart object recommendation system, wherein, social relationships in terms of SIoT knowledge graph and social similarity together with collaborative filtering model is amalgamated to recommend an object in SIoT application. Likewise in \cite{8662641}, an experience and reputation trust evaluation model is delineated to recruit the mobile nodes for mobile crowdsourcing. The experimental evaluation is performed on a real-world dataset and the effectiveness of the same is validated by comparing the model with the state-of-the-art schemes.   

Numerous studies in the literature also suggest the idea of utilizing various techniques such as a fuzzy logic-based model \cite{8737491}, machine learning-based schemes \cite{8364607}\cite{9148767}, and regression model \cite{Wang2014LogitTrustA} to ascertain the single trust score as the traditional weighted sum method has many drawbacks. Wang {\etal} \cite{Wang2014LogitTrustA} presented a logit regression-based model to estimate the trust which is based on the conditional probability of a credible service by a service provider. Nevertheless, this technique requires more observation to deal with different recommendation-based attacks. Xin {\etal} \cite{8737491} designed a context-aware fuzzy logic-based trust model to build a reliable trustworthy relationship among the objects, wherein, centrality and community-interest are considered to quantify the trust of an object. Also, fuzzy logic-based inference rules are formed to synthesize the selected trust parameters in order to ascertain a single trust score. Finally, one of the most recent works related to our research was carried out by Upul {\etal} \cite{8364607}. The authors staged a machine learning-based trust framework model based on the social profile of a node, wherein different social features are accumulated by exploiting machine learning-based to ascertain the direct trust metric of any node in the IoT network. However, their approach lacks the idea of incorporating recommendations as indirect trust along with the direct trust observation.   

\section{Proposed Trust Evaluation Model}

This section formally defines the problem focused in this paper, and subsequently introduces the proposed trust evaluation model.  

\subsection{Problem Statement}

In order to quantify the trustworthiness of an object in the SIoT environment, we use social information (i.e., Friends, Interest Group, Working Relationships, etc) of an object, and $object-object$ interactions as our primary data source. The social information is the collection of object's friends ($F$), object's social interest communities ($C$), and co-work relationship ($CW$) information in terms of mutlicast interactions, whereas $object-object$ interactions gives the insight of cooperativeness ($CoP$) in the form of $successful$ and $unsuccessful$ interactions among the participating objects. The data sources are defined as the set of triplets $\{ F, C, CW \}$, and $object-object$ interactions as cooperativeness ($CoP$). 

Let $O = \{o_1,o_2,...,o_n\}$ represents the set of objects and given the set of triplets $\{ F, C, CW \}$ for each object, and the $object-object$ interactions, the targeted problem of this paper can be formulated as quantifying the trust of an object $o_i$ towards another object $o_j$ at any time interval $t$ by using the given set of triplets and cooperativeness, and is expressed as the composition of all the trust parameters as follows:   

\begin{equation}
     Trust_t(o_i,o_j) \ = \ <F, C, CW, CoP>  
\end{equation}

\begin{figure*}[ht]
\centering
\includegraphics[width=0.9\linewidth]{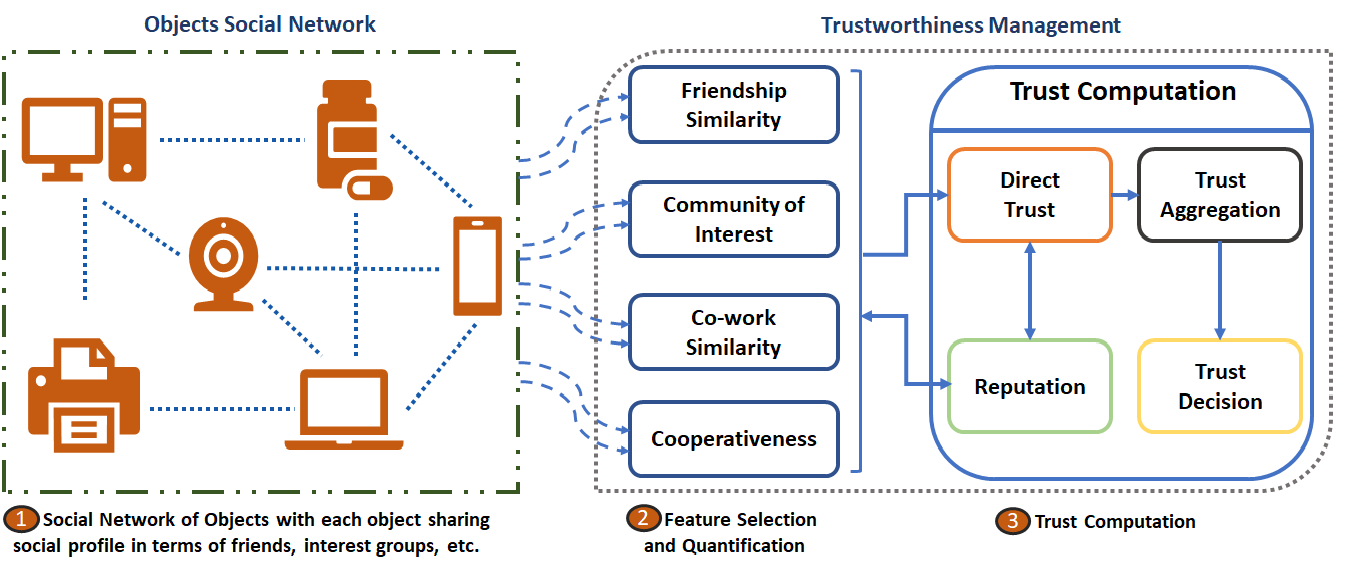}
\caption{Proposed SIoT-based trust evaluation model} \label{fig:trust_model}
\end{figure*}

\subsection{Trust Evaluation}

For the purpose of quantifying trustworthiness of the object from their social information and interactions, we propose a computational model depicted in Figure \ref{fig:trust_model} which considers all of the salient aspects of trustworthiness management. As can be seen in figure, the first step to compute the trustworthiness is to get the social information of trustee, subsequently the next step is the quantify the selected feature by using the information provided in first step. Feature extraction component addresses the issue of selecting and quantifying the trust metrics suitable for a particular Social IoT application, and the designated metrics for this paper are $Friendship$ $Similarity$, $Community-of-Interest$, $Co-work$  $Similarity$, and $Cooperativeness$. After that, trustworthiness of an object is computed using both the direct trust and the reputation, then the trust aggregation process is employed to accumulate the independent social trust metrics to form the final trust score via employing a machine learning-driven scheme. Finally, the trust decision component provides the information on whether a node is trustworthy or untrustworthy, and is denoted as $Trust_{t} (o_i,o_j)$ at time $t$ between any two objects $o_i$ as trustor and $o_j$ as trustee.

\subsubsection{Friendship Similarity ($T_{FS}^t$)}

This trust feature signifies the importance of an object $o_j$ vis-à-vis the social relationship of the object $o_i$ locally among its immediate neighbours at any time $t$. Besides, friendship similarity prohibits the malicious objects from establishing counterfeit social relationships to get the advantage of higher similarity. It is widely accepted that friends are slanted to cooperate with each other and therefore, highly similar objects can be selected for service discovery and provisioning or a common task. An object can describe the friendship of a neighbouring object as follows:

\begin{equation}
    T_{FS}^t (o_i,o_j) = \frac{|F_{o_i} \cap F_{o_j}|}{|F_{o_i} \cup F_{o_j}|}
\end{equation}
wherein, $F_{o_i}$ and $F_{o_j}$ refers to a set of friends of object $o_i$ and object $o_j$ respectively. Furthermore, $|.|$ shows the cardinality of a set which gives the count on the number of elements in the set.

% \begin{figure*}[ht]
% \begin{center}
%   \subfloat[CoI vs FS]{
% 	\begin{minipage}[c][]{
% 	   0.3\textwidth}
% 	   \centering
% 	   \includegraphics[width=\linewidth]{coi_fs_cls.png}
% 	\end{minipage}}
%  \hfill 	
%   \subfloat[CoI vs CWS]{
% 	\begin{minipage}[c][]{
% 	   0.3\textwidth}
% 	   \centering
% 	   \includegraphics[width=\linewidth]{coi_cwr_cls.png}
% 	\end{minipage}}
%  \hfill	
%   \subfloat[CoI vs CoP]{
% 	\begin{minipage}[c][]{
% 	   0.3\textwidth}
% 	   \centering
% 	   \includegraphics[width=\linewidth]{coi_cop_cls.png}
% 	\end{minipage}}
% \caption{Clustering on different pairs of trust features}
% \end{center}
% \end{figure*}
\begin{figure*}[h]
\centering
\includegraphics[width=\linewidth]{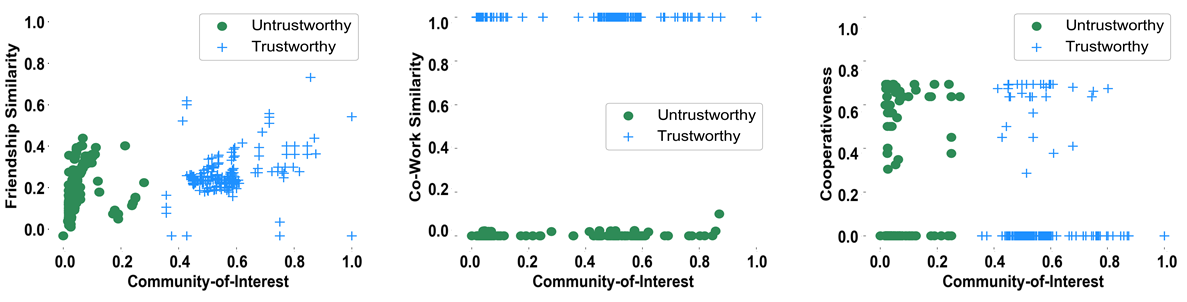}
\caption{Clustering on different pairs of trust features} \label{fig:clustering_results}
\end{figure*}

\begin{figure*}[h]
\centering
\includegraphics[width=\linewidth]{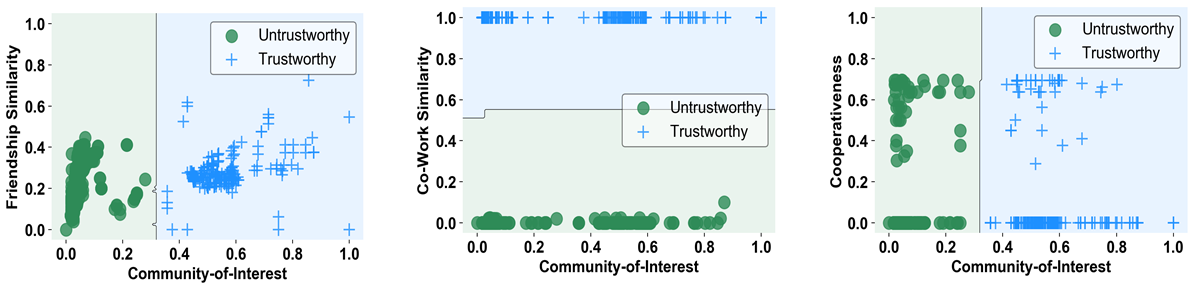}
\caption{Classification on different pairs of trust features} \label{fig:classification_results}
\end{figure*}

% \begin{figure*}[ht]
%   \subfloat[CoI vs FS]{
% 	\begin{minipage}[c][]{
% 	   0.3\textwidth}
% 	   \centering
% 	   \includegraphics[width=\linewidth]{coi_fs_clf.png}
% 	\end{minipage}}
%  \hfill 	
%   \subfloat[CoI vs CWS]{
% 	\begin{minipage}[c][]{
% 	   0.3\textwidth}
% 	   \centering
% 	   \includegraphics[width=\linewidth]{coi_cwr_clf.png}
% 	\end{minipage}}
%  \hfill	
%   \subfloat[CoI vs CoP]{
% 	\begin{minipage}[c][]{
% 	   0.3\textwidth}
% 	   \centering
% 	   \includegraphics[width=\linewidth]{coi_cop_clf.png}
% 	\end{minipage}}
% \caption{Classification on different pairs of trust features}
% \end{figure*}

\subsubsection{Community-of-Interest ($T_{CoI}^t$)}
This property facilitates in computing the community-based trust feature of a trustee $o_j$ vis-à-vis the trustor $o_i$ at time $t$, wherein both the objects share common interest groups like social groups, e-commerce, and so forth, i.e., it is an indication of the common interest between them. In SIoT environment, object collaborate with at least one interest group, and two objects have a greater chance to build up close contact with one another if they have a high level of community of interest \cite{6513398}. In contrast to the example of friendship similarity, the community of interest introduced in objects do not change instinctively, thus, each object needs to store a list of its owner's interest group. Mathematically, $T_{CoI}^t$ is computed as follows:

\begin{equation}
    T_{CoI}^t (o_i,o_j) = \frac{|C_{o_i} \cap C_{o_j}|}{|C_{o_i} \cup C_{o_j}|}
    \label{eq:coi}
\end{equation}
here, $C_{o_i}$ and $C_{o_j}$ depict the interest group of objects $i$ and $j$ respectively. The higher the degree of common interest (i.e. $T_{CoI}^t (o_i,o_j)$), the more prominent is the similarity between the objects.

% By Utilizing Eq. \ref{eq:coi}, the degree of common interests between a trustor and a trustee is computed. 

\subsubsection{Co-work Similarity ($T_{CWS}^t$)}

Co-work similarity gives the notion of trust when two or more objects collaborate with each other to accomplish a common goal. In this type of trust, more focus is on the working relation instead of their physical closeness. Precisely, $T_{CWS}^t$ score is measured as the ratio of common multicast interaction among the object to the total number of multicast interactions at any time $t$, and as given in Eq. \ref{eq:cws} 
. 
\begin{equation}
    T_{CWS}^t (o_i,o_j) = \frac{|M_{o_i} \cap M_{o_j}|}{|M_{o_i} \cup M_{o_j}|}
    \label{eq:cws}
\end{equation}
wherein, $M_{o_i}$ represents the multicast interactions of object $o_i$, whereas, $M_{o_j}$ symbolizes the multicast interactions of object $o_j$. 

\begin{figure*}[ht]
  \subfloat[Time: 4 Hours]{
	\begin{minipage}[c][]{
	   0.3\textwidth}
	   \centering
	   \includegraphics[width=\linewidth]{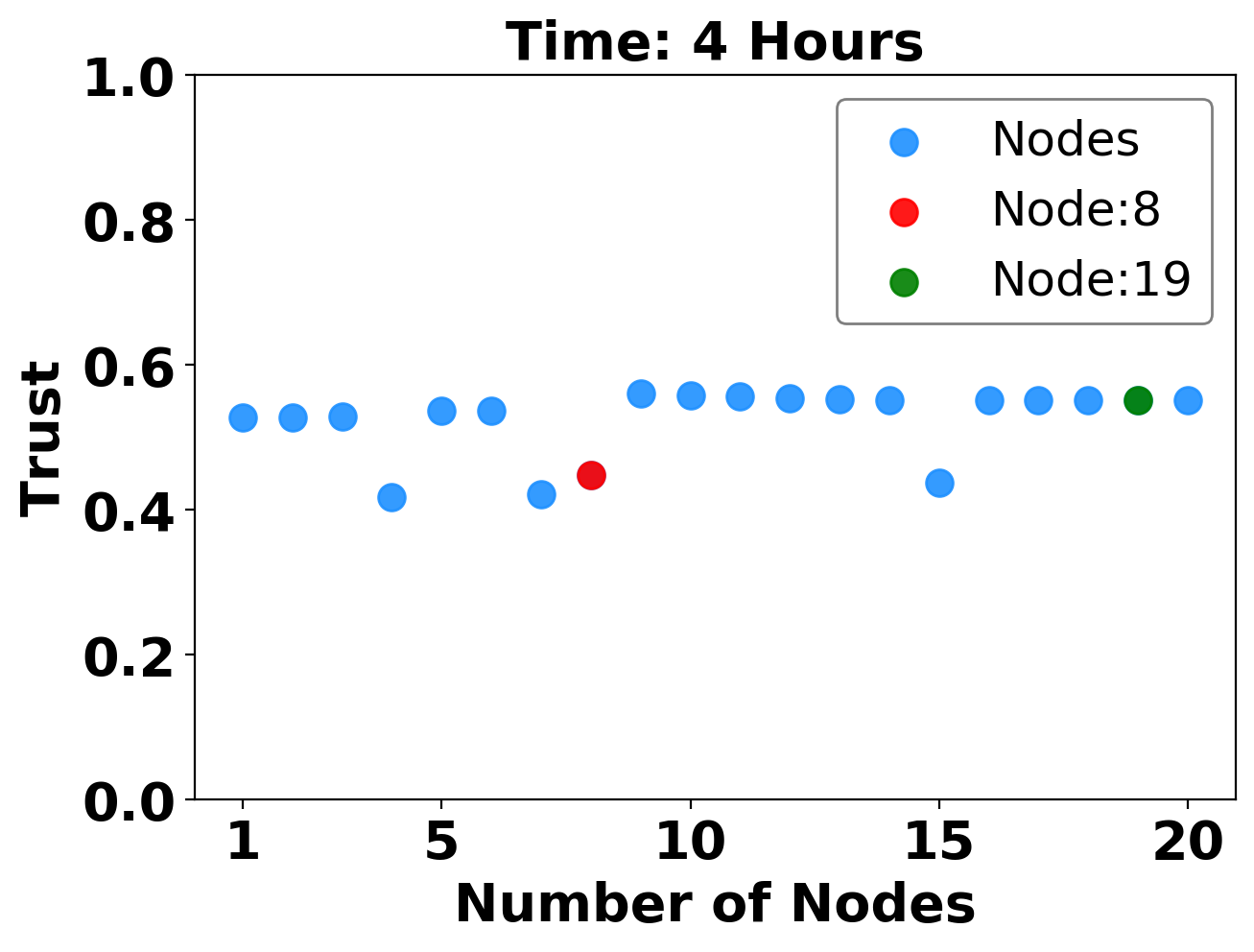}
	\end{minipage}}
 \hfill 	
  \subfloat[Time: 8 Hours]{
	\begin{minipage}[c][]{
	   0.3\textwidth}
	   \centering
	   \includegraphics[width=\linewidth]{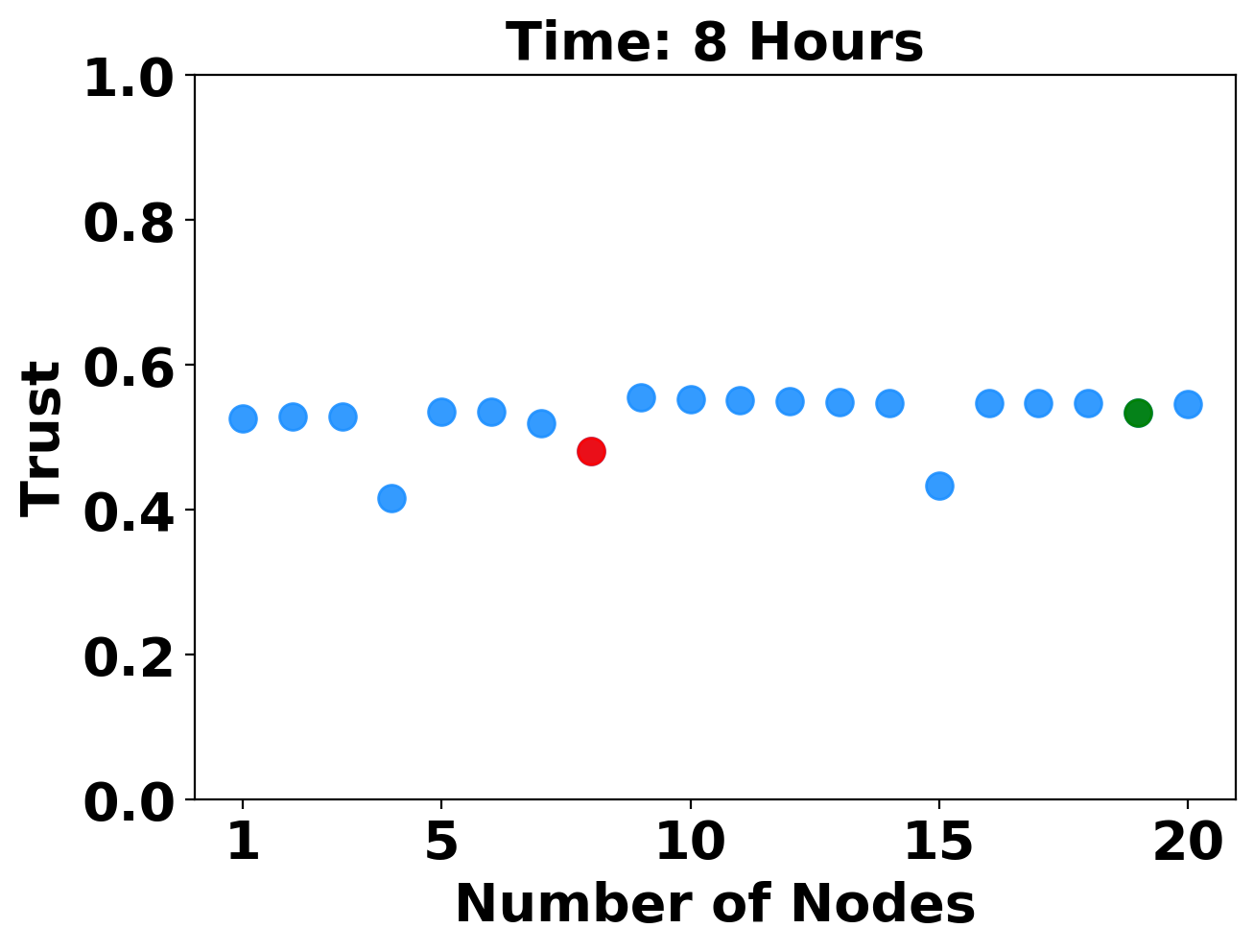}
	\end{minipage}}
 \hfill	
  \subfloat[Time: 12 Hours]{
	\begin{minipage}[c][]{
	   0.3\textwidth}
	   \centering
	   \includegraphics[width=\linewidth]{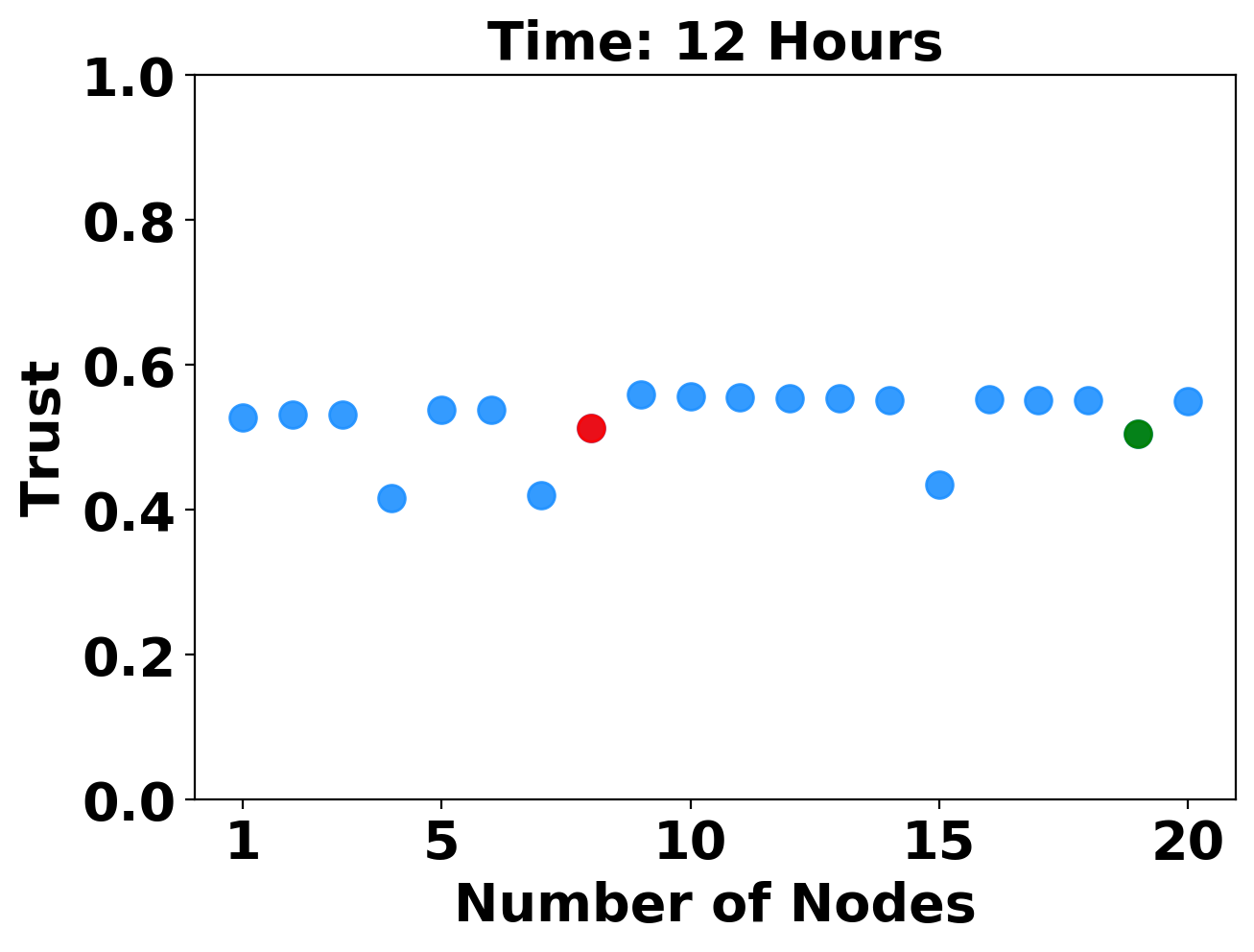}
	\end{minipage}}
	 \hfill	
  \subfloat[Time: 16 Hours]{
	\begin{minipage}[c][]{
	   0.3\textwidth}
	   \centering
	   \includegraphics[width=\linewidth]{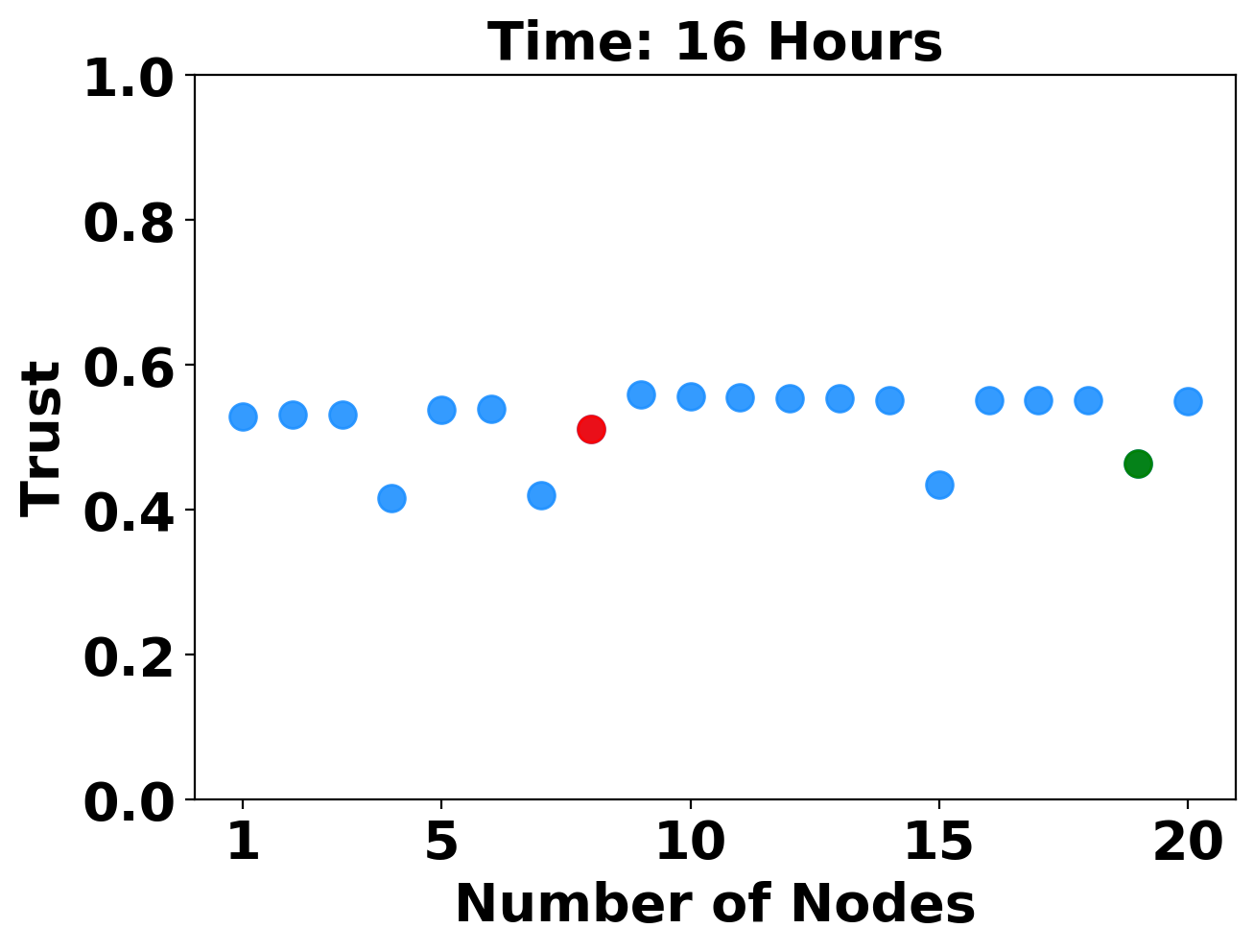}
	\end{minipage}}
	 \hfill	
  \subfloat[Time: 20 Hours]{
	\begin{minipage}[c][]{
	   0.3\textwidth}
	   \centering
	   \includegraphics[width=\linewidth]{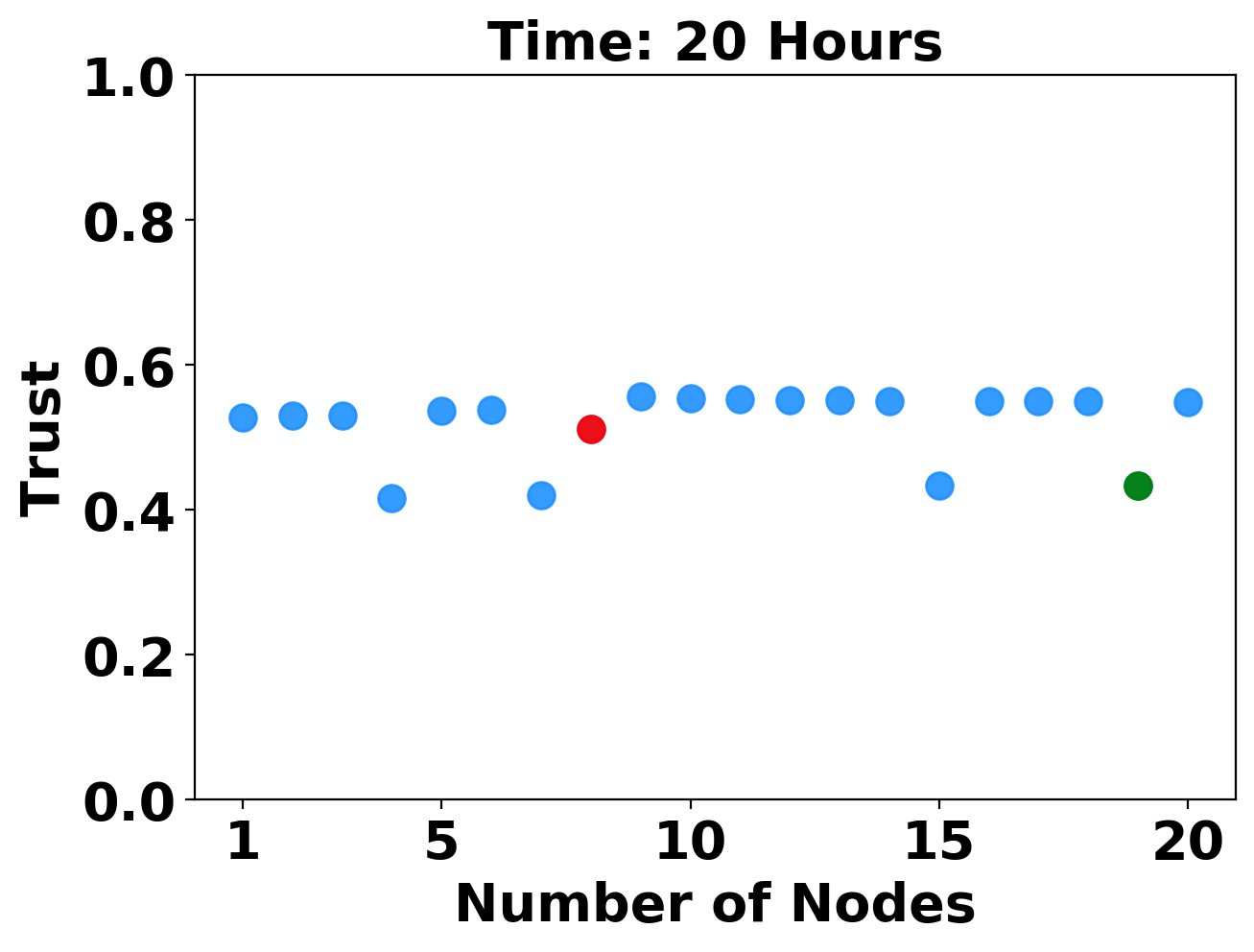}
	\end{minipage}}
	 \hfill	
  \subfloat[Time: 24 Hours]{
	\begin{minipage}[c][]{
	   0.3\textwidth}
	   \centering
	   \includegraphics[width=\linewidth]{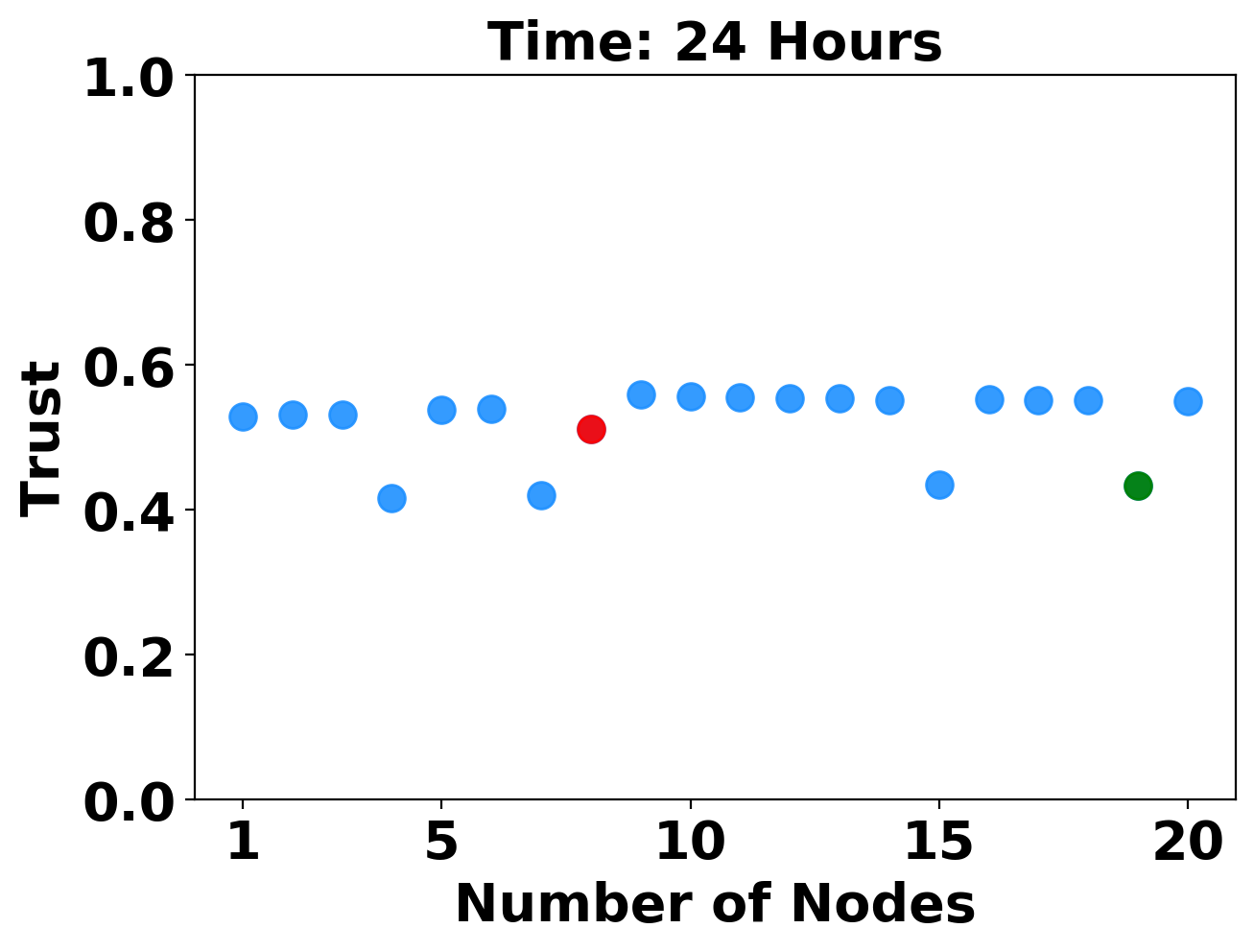}
	\end{minipage}}
\caption{Trust score of 20 Nodes over a period of 24 hours}
\label{fig:trust_time}
\end{figure*}

\subsubsection{Cooperativeness ($T_{CoP}^t$)} 

Cooperativeness manifests whether a trustee is socially cooperative in terms of interactions with the trustor or not, i.e., an object can behave maliciously for a specific service and try to manipulate the authenticity of the information for the other service, thus CoP maintains the content consistency and offers trustworthy service. Since CoP refers to a measure of balance in the interaction between the objects, so we can utilize the concept of entropy function delineated in \cite{5484757} to get CoP-based trust feature as:

\begin{equation}
    T_{CoP}^t (o_i,o_j) = -T_p \ log (T_p) - (1-T_p) \ log (1-T_p) \label{eq:cop}   
\end{equation}
where, $T_p$ represents fraction of messages during the interaction between the object $o_i$ vis-à-vis object $o_j$, and is computed as follows:

\begin{equation}
     T_p = \frac{Successful\_Interaction}{Total\_Interaction}
\end{equation}
% where $Successful\_Interaction$ represents the number of successful packets transmitted at any time $t$ during the interaction between object $i$ and object $j$ while $Total\_Interaction$ gives the count of the total number of packets transmitted.

Decisively, to estimate the single trust score, a notable methodology to combine all trust features is to use a conventional weighted sum method as shown in Eq. \eqref{eq:total} while determining the appropriate value of weights.  

\begin{dmath}
Trust_{t} (o_i,o_j) = \sum_{i=1}^{n} w_i T_{X_{o_i}}^t(o_i,o_j) \label{eq:total}
\end{dmath}
where $X$ represents the trust features (i.e., $CoI$, $FS$, $CWS$ and $CoP$), $w$ symbolize the weight of each trust feature, and $n$ gives the count on the total number of trust features used.

Nevertheless, the weighted sum approach has numerous disadvantages including but not limited to an unending number of conceivable outcomes with regards to assessing a weighting factor for each feature and inability to recognize which trust feature makes the most effect on the trust in a specific context. Therefore, to overcome these drawback, we hereby propose a machine learning-based scheme which combines all trust features to ascertain an overall trust value, and to identify the impact of each feature on the final trust decision.

\section{Simulation Setup}
To get the statistical estimate for the trust features mentioned in Section 3.2, we have utilized the CRAWDED dataset from SIGCOMM-2009 \cite{thlab-sigcomm2009-20120715} to map these traces of data in the form, to be exploited for SIoT environment. This dataset comprises of $76$ nodes with $18,226$ interaction between them for over a period of four days. These traces contain the social information of nodes (i.e., friendship, interested communities, interactions, and message logs). There are $5,776$ pairs of interactions for these $76$ nodes, and trust features for each pair is computed with at least one interaction between these pairs and these features are formulated in the form of feature matrix given in Eq. \ref{eq:fm}. To assess the proposed model, we have detached the data into hours and have checked the trust score of arbitrarily chosen $20$ nodes over a time of $24$ hours. 

%since as it is not feasible to visualize all of the features altogether.  

\begin{dmath}
Feature \ Matrix = \begin{bmatrix}
$CoI$_1 & $FS$_1 & $CWS$_1 & $CoP$_1\\
\vdots & \vdots & \vdots & \vdots\\
$CoI$_m & $FS$_m & $CWS$_m & $CoP$_m
\end{bmatrix} \label{eq:fm}
\end{dmath}
where $m$ shows the number of interactions.

\subsection{Machine Learning-driven Aggregation Scheme}

To overcome the shortcoming of the weighted sum method as discussed earlier, a machine learning-driven (ML-driven) aggregation scheme is proposed to accumulate the extracted features. So as to accomplish this, we first utilize an unsupervised learning algorithm (i.e., k-means clustering) to distinguish two distinct classes, in particular trustworthy and untrustworthy \cite{Breiman2001}. The primary motivation to utilize unsupervised learning over supervised learning is because of the unavailability of the labeled training set. The k-means algorithm compels two initial information sources; initial centroid points ($C$) and the number of clusters ($C_k$). In the proposed model, we have randomly assigned the initial centroids for two clusters, namely trustworthy and trustworthy, wherein, the objects close to 1 are marked as trustworthy and objects with values near the origin $(0,0)$ are marked as untrustworthy.

Subsequently, a multi-class classification algorithm like random forest \cite{Amatriain2011} is utilized to train the model, and to identify the best decision boundary that distinguish trustworthy and untrustworthy interactions.  There are a few motives to employ random forest for classification. Random forest is appropriate to recognize the most significant features in the dataset as well as the weights of each trust feature to acquire the single trust score. Furthermore, this classification algorithm makes use of multiple random decision trees on the same dataset to avoid overfitting.  
Finally, for demonstrations purpose, we just consider the pairs of trust features for both the clustering and the classification as opposed to utilizing all the features at once.

\begin{figure*}[ht]
  \subfloat[CoI, Trust vs Time]{
	\begin{minipage}[c][]{
	   0.33\linewidth}
	   \centering
	   \includegraphics[width=\linewidth]{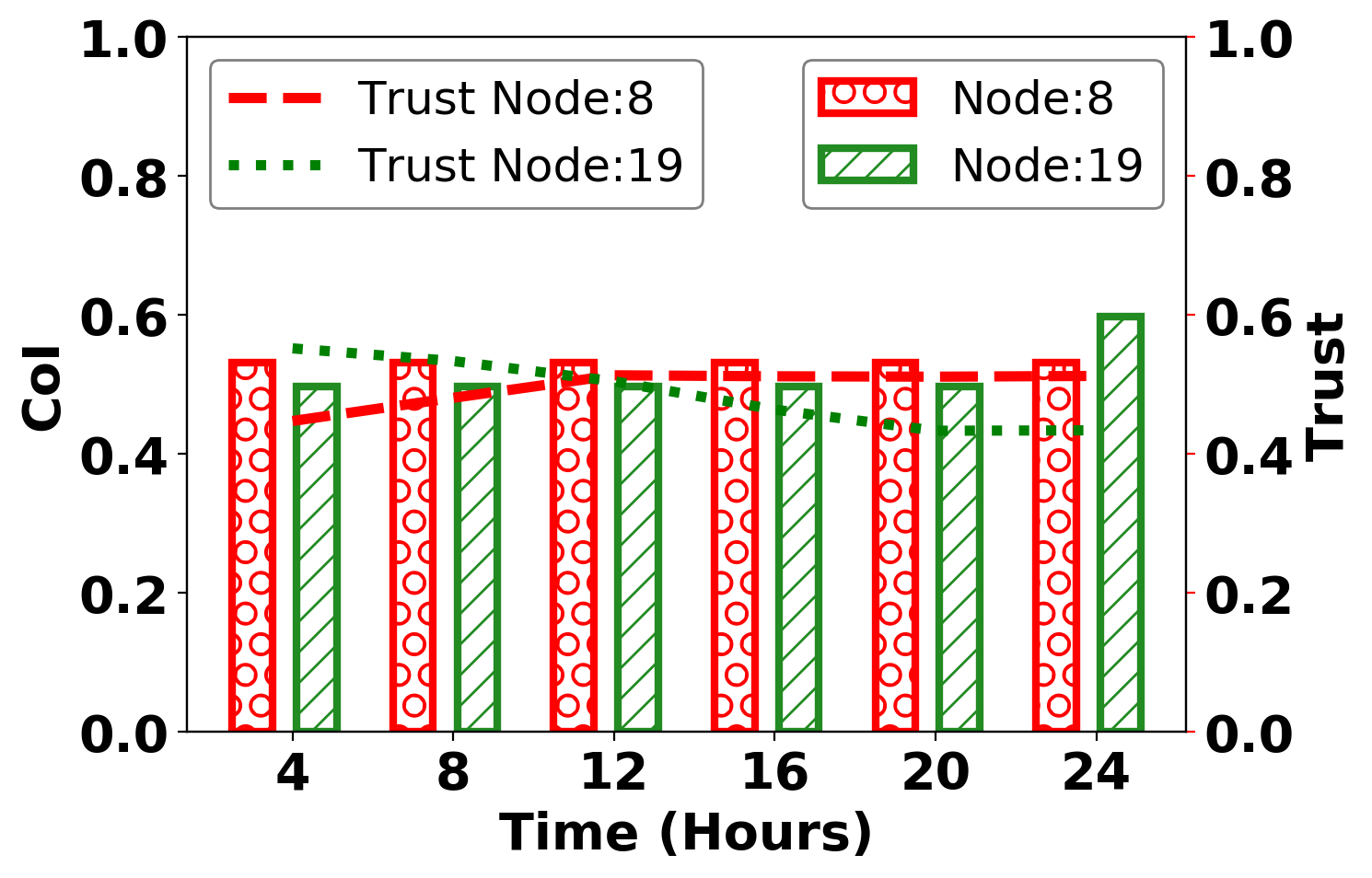}
	\end{minipage}}
 \hspace{1em}	
  \subfloat[FS, Trust vs Time]{
	\begin{minipage}[c][]{
	   0.33\linewidth}
	   \centering
	   \includegraphics[width=\linewidth]{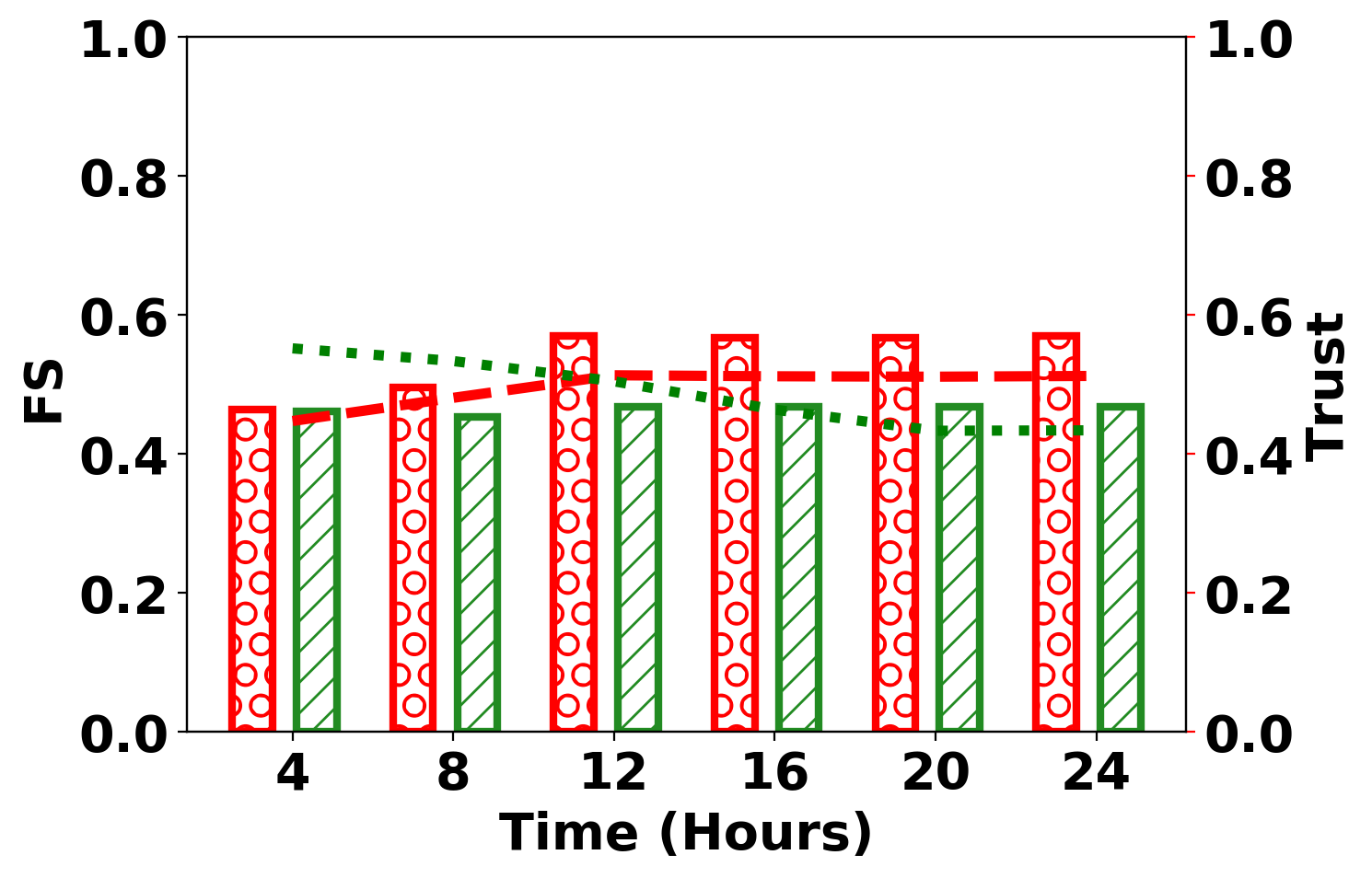}
	\end{minipage}}
	\hspace{1em}
  \subfloat[CWS, Trust vs Time]{
	\begin{minipage}[c][]{
	   0.33\linewidth}
	   \centering
	   \includegraphics[width=\linewidth]{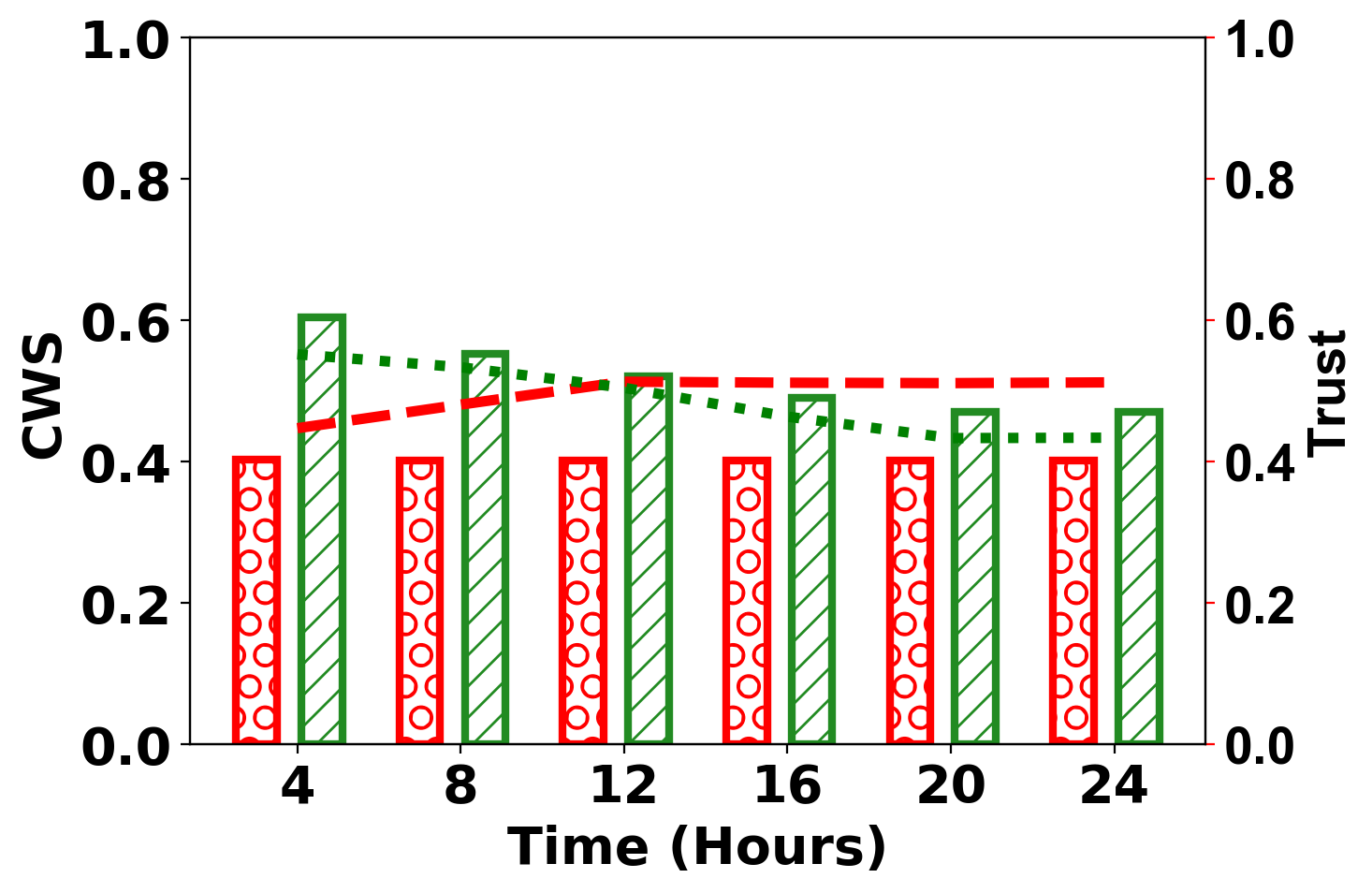}
	\end{minipage}}
	 \hspace{1em}
	\subfloat[CoP, Trust vs Time]{
		\begin{minipage}[c][]{
	   0.33\linewidth}
	   \centering
	   \includegraphics[width=\linewidth]{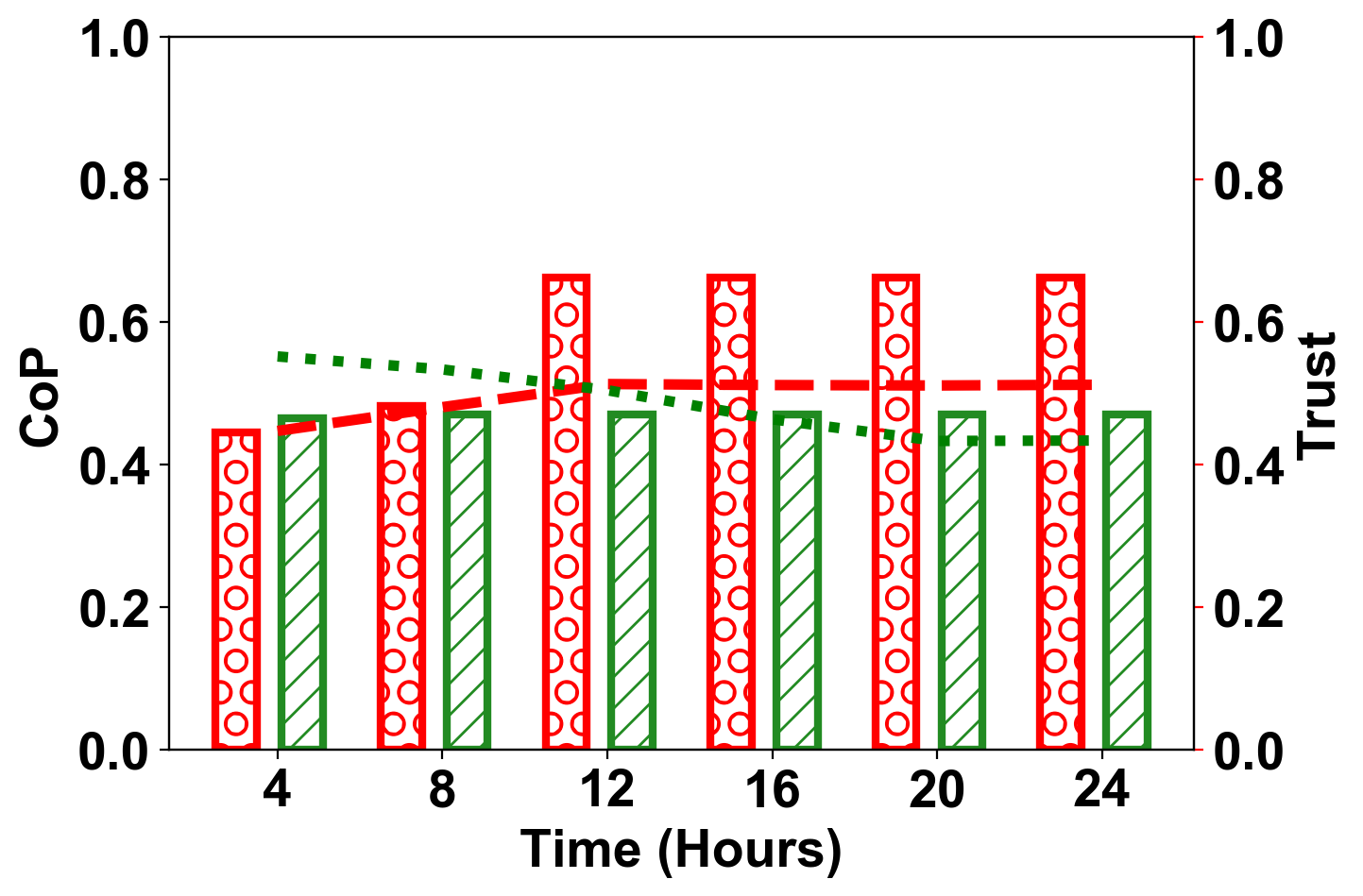}
	\end{minipage}}
\caption{Effect of CoI, FS, CWS, and CoP score on nodes trust}
\label{fig:effect_trust}
\end{figure*}

% \begin{figure*}[h]
% \centering
% \includegraphics[width=0.7\linewidth]{nodes_features_trust.png}
% \caption{Effect of CoI, FS, CWS, and CoP score on nodes trust} \label{fig:effect_trust}
% \end{figure*}

\vspace{-0.75em}

\section{Results and Discussion}

This section describes the performance evaluation of the proposed model deliberated in Section 3. As discussed, we merely consider pairs of trust features for both visualization and distribution of trust via comparing $CoI$ with $FS$, $CWS$, and $CoP$ respectively. As portrayed in Figure \ref{fig:clustering_results}, it can be clearly observed how the clustering algorithm successfully classifies the interaction as trustworthy or untrustworthy. Subsequently, as depicted in Figure \ref{fig:classification_results}, the classification algorithm (Random Forest) clearly identifies the pair-wise decision boundary between the interactions and is capable of classifying the futuristic interactions as trustworthy or untrustworthy. Additionally, the classification algorithm also manifests the weightage (i.e., feature importance) of each individual feature so as to ascertain a single trust score, wherein the weightage of each feature \{$CoI$, $FS$, $CWS$, and $CoP$\} is \{$23.11\%$, $12.96\%$, $60.75\%$, and $3.16\%$\}. Besides, we have evaluated the trust model in terms of accuracy and considered the following three evaluation metrics: 

\begin{itemize}
    \item[1.] \emph{Precision:} Precision is defined as the accuracy of a model to classify trustworthy objects as trustworthy and untrustworthy objects as untrustworthy. It is measured as the ratio of correctly predicted untrustworthy observations to the total number of untrustworthy observations (as per untrustworthy class). The value of precision is class dependent.   
        % \begin{equation}
        %   Precision = \frac{TP}{TP+FP}
        % \end{equation}
        % wherein, $TP$ represents trust positive which means observation of both actual and predicted class are same, whereas, $FP$ represents false positive which means a trustworthy object predicted as untrustworthy (as per untrustworthy class).   
    \item[2.] \emph{Recall:} Recall is also class dependent and is referred to the proportion of trustworthy or untrustworthy objects that have been correctly identified. It is measured as the ratio of correctly predicted positive (i.e., trustworthy or untrustworthy) observation to all the observations in the actual class. 
        % \begin{equation}
        %   Recall = \frac{TP}{TP+FN}
        % \end{equation}
        % here, $FN$ represents false negative which means a untrustworthy object predicted as trustworthy as per untrustworthy class. 
    \item[3.] \emph{F1-Score:} It is used to measure the accuracy of a model by considering weighted average of precision and recall.
        % \begin{equation}
        %   F1-Score = 2* \frac{Precision*recall}{Precision+Recall}
        % \end{equation}
\end{itemize}

Overall, Table 1 gives the score of each of the above evaluation metrics for both the classes, \emph{trustworthy (T)} and \emph{untrustworthy (U)}, and it can observed that our model gives the high precision and recall values with an accuracy of $98.8\%$.   

Furthermore, Figure \ref{fig:trust_time} depicts the trust score of randomly selected $20$ nodes over a period of $24$ hours. As can be seen from the figure, the trust result of nearly all the nodes remains the same. However, trust of two nodes ($Node 8 \ and \  Node 19$) varies with time, i.e., trust score of $Node 8$ increases with time and turns out to be stable after time $t=12 \ hours$. Besides, the trust score of $Node 19$ decreases with time, i.e., starting from the trust score of almost $0.6$ at time $t=4 \  hours$ it drops down to $0.4$ at the end. Subsequently, Figure \ref{fig:effect_trust} justifies the variation in the trust score of two nodes ($Node 8 \ and \  Node 19$) and illustrates the effect of each trust feature on trust of each node. It can be observed from Figure \ref{fig:effect_trust}(a) and Figure \ref{fig:effect_trust}(c) that the CoI and CWS score for Node: 8 remains the same but the change in score of FS and CoP prompt the increase in the trust result of Node:8. Similarly, the decrease in the CWS score as shown in Figure \ref{fig:effect_trust}(c) decreases the trust score of Node:19 as all other trust features remain the same. Overall, the computed score for each of the trust parameters for 20 nodes over the period of time is depicted in Table 2.     

\begin{table}[h]
    \centering
        \caption{Performance Evaluation}
\begin{tabular}{|c|c|c|c|}
 \hline
 \multicolumn{4}{|c|}{\textbf{Accuracy = $98.8\%$}} \\
 \hline
 \hline
 \textbf{Classes} & \textbf{Precision} & \textbf{Recall} & \textbf{F-Score}\\
 \hline
 Untrustworthy   & 1.0 & 0.97 & 0.99\\
 Trustworthy & 0.98  & 1.00   & 0.99\\
 \hline
\end{tabular}
    \label{tab:my_label}
\end{table}

\section{Conclusion}
In this paper, we have envisaged an efficient time-aware trust evaluation model to identify untrustworthy objects in the SIoT network. Precisely, the proposed trust model consider SIoT relationships in terms of friendship, community-of-interest, co-work similarity, and cooperativeness as the trust parameters. Furthermore, a machine learning-driven aggregation scheme is introduced so as to synthesize these trust parameters to ascertain a \emph{single} trust score. Simulation results demonstrate the effectiveness of the model via segregating trustworthy and untrustworthy objects and further provides the variation in the trust score of each node over a period of time.

In order to further develop our idea, we intend to verify the convergence and resilience property of the proposed model by incorporating context-awareness (i.e., environment conditions, energy, and time) in a dynamically changing SIoT environment. Additionally, quantifying and employing social strength \cite{8463599} among the objects having social connection as a trust parameter would result in the precise determination of trust. 

% \section{Acknowledgement}
% The corresponding author sincerely acknowledges the generous support of the Higher Education Commission of Pakistan and Macquarie University for funding the research at hand via its ‘Macquarie University Research Excellence Award (Allocation No. 2019050)’.

\begin{table*}[h]
%\scriptsize
\fontsize{5.35}{7.25}\selectfont
\captionsetup[table]{skip=0pt}  
\caption{Trust Parameters Data for 20 Nodes}
{\renewcommand{\arraystretch}{1.9} 
\begin{tabular}{c|cccc|cccc|cccc|cccc|cccc|cccc}
\hline
 & \multicolumn{4}{c|}{\textbf{Time: 1}} & \multicolumn{4}{c|}{\textbf{Time:2}} & \multicolumn{4}{c|}{\textbf{Time: 3}} & \multicolumn{4}{c|}{\textbf{Time: 4}} & \multicolumn{4}{c|}{\textbf{Time: 5}} & \multicolumn{4}{c}{\textbf{Time: 6}} \\ \hline
\textbf{Id} & \textbf{CoI} & \textbf{FS} & \textbf{CWS} & \textbf{CoP} & \textbf{CoI} & \textbf{FS} & \textbf{CWS} & \textbf{CoP} & \textbf{CoI} & \textbf{FS} & \textbf{CWS} & \textbf{CoP} & \textbf{CoI} & \textbf{FS} & \textbf{CWS} & \textbf{CoP} & \textbf{CoI} & \textbf{FS} & \textbf{CWS} & \textbf{CoP} & \textbf{CoI} & \textbf{FS} & \textbf{CWS} & \textbf{CoP} \\ \hline
\textbf{1} & 0.43 & 0.44 & 0.60 & 0.43 & 0.43 & 0.43 & 0.59 & 0.43 & 0.43 & 0.43 & 0.59 & 0.43 & 0.43 & 0.43 & 0.59 & 0.44 & 0.43 & 0.43 & 0.59 & 0.44 & 0.43 & 0.43 & 0.59 & 0.44 \\ \hline
\textbf{2} & 0.43 & 0.44 & 0.60 & 0.40 & 0.43 & 0.44 & 0.60 & 0.40 & 0.43 & 0.45 & 0.60 & 0.42 & 0.43 & 0.45 & 0.60 & 0.42 & 0.43 & 0.45 & 0.60 & 0.42 & 0.43 & 0.45 & 0.60 & 0.42 \\ \hline
\textbf{3} & 0.43 & 0.44 & 0.60 & 0.41 & 0.43 & 0.44 & 0.60 & 0.41 & 0.43 & 0.45 & 0.60 & 0.42 & 0.43 & 0.45 & 0.60 & 0.42 & 0.43 & 0.45 & 0.60 & 0.42 & 0.43 & 0.45 & 0.60 & 0.42 \\ \hline
\textbf{4} & 0.43 & 0.46 & 0.40 & 0.42 & 0.43 & 0.46 & 0.40 & 0.42 & 0.43 & 0.46 & 0.40 & 0.42 & 0.43 & 0.46 & 0.40 & 0.42 & 0.43 & 0.46 & 0.40 & 0.42 & 0.43 & 0.46 & 0.40 & 0.42 \\ \hline
\textbf{5} & 0.44 & 0.47 & 0.60 & 0.42 & 0.44 & 0.47 & 0.60 & 0.42 & 0.44 & 0.47 & 0.60 & 0.42 & 0.44 & 0.47 & 0.60 & 0.42 & 0.44 & 0.47 & 0.60 & 0.42 & 0.44 & 0.47 & 0.60 & 0.42 \\ \hline
\textbf{6} & 0.44 & 0.47 & 0.60 & 0.44 & 0.44 & 0.46 & 0.60 & 0.45 & 0.44 & 0.47 & 0.60 & 0.45 & 0.44 & 0.47 & 0.60 & 0.45 & 0.44 & 0.47 & 0.60 & 0.45 & 0.44 & 0.47 & 0.60 & 0.45 \\ \hline
\textbf{7} & 0.44 & 0.47 & 0.40 & 0.44 & 0.44 & 0.46 & 0.40 & 0.45 & 0.44 & 0.47 & 0.40 & 0.45 & 0.44 & 0.47 & 0.40 & 0.45 & 0.44 & 0.47 & 0.40 & 0.45 & 0.44 & 0.47 & 0.40 & 0.45 \\  \hline
\color{black}\textbf{8} & \color{black}\textbf{0.53} & \color{black}\textbf{0.46} & \color{black}\textbf{0.40} & \color{black}\textbf{0.45} & \color{black}\textbf{0.53} & \color{black}\textbf{0.50} & \color{black}\textbf{0.40} & \color{black}\textbf{0.48} & \color{black}\textbf{0.53} & \color{black}\textbf{0.57} & \color{black}\textbf{0.40} & \color{black}\textbf{0.66} & \color{black}\textbf{0.53} & \color{black}\textbf{0.57} & \color{black}\textbf{0.40} & \color{black}\textbf{0.66} & \color{black}\textbf{0.53} & \color{black}\textbf{0.57} & \color{black}\textbf{0.40} & \color{black}\textbf{0.66} & \color{black}\textbf{0.53} & \color{black}\textbf{0.57} & \color{black}\textbf{0.40} & \color{black}\textbf{0.66} \\ \hline
\textbf{9} & 0.53 & 0.46 & 0.60 & 0.46 & 0.53 & 0.45 & 0.60 & 0.46 & 0.53 & 0.47 & 0.60 & 0.46 & 0.53 & 0.47 & 0.60 & 0.46 & 0.53 & 0.47 & 0.60 & 0.46 & 0.53 & 0.47 & 0.60 & 0.46 \\ \hline
\textbf{10} & 0.52 & 0.46 & 0.60 & 0.41 & 0.52 & 0.45 & 0.60 & 0.41 & 0.52 & 0.47 & 0.60 & 0.41 & 0.52 & 0.47 & 0.60 & 0.41 & 0.52 & 0.47 & 0.60 & 0.41 & 0.52 & 0.47 & 0.60 & 0.41 \\ \hline
\textbf{11} & 0.52 & 0.46 & 0.60 & 0.40 & 0.52 & 0.45 & 0.60 & 0.40 & 0.52 & 0.47 & 0.60 & 0.40 & 0.52 & 0.47 & 0.60 & 0.40 & 0.52 & 0.47 & 0.60 & 0.40 & 0.52 & 0.47 & 0.60 & 0.40 \\ \hline
\textbf{12} & 0.51 & 0.46 & 0.60 & 0.44 & 0.51 & 0.45 & 0.60 & 0.44 & 0.51 & 0.47 & 0.60 & 0.44 & 0.51 & 0.47 & 0.60 & 0.44 & 0.51 & 0.47 & 0.60 & 0.44 & 0.51 & 0.47 & 0.60 & 0.44 \\ \hline
\textbf{13} & 0.50 & 0.46 & 0.60 & 0.40 & 0.50 & 0.45 & 0.60 & 0.45 & 0.50 & 0.47 & 0.60 & 0.47 & 0.50 & 0.47 & 0.60 & 0.47 & 0.50 & 0.47 & 0.60 & 0.47 & 0.50 & 0.47 & 0.60 & 0.47 \\ \hline
\textbf{14} & 0.50 & 0.46 & 0.60 & 0.41 & 0.50 & 0.45 & 0.60 & 0.41 & 0.50 & 0.47 & 0.60 & 0.41 & 0.50 & 0.47 & 0.60 & 0.41 & 0.50 & 0.47 & 0.60 & 0.41 & 0.50 & 0.47 & 0.60 & 0.41 \\ \hline
\textbf{15} & 0.50 & 0.46 & 0.40 & 0.43 & 0.50 & 0.45 & 0.40 & 0.45 & 0.50 & 0.47 & 0.40 & 0.45 & 0.50 & 0.47 & 0.40 & 0.45 & 0.50 & 0.47 & 0.40 & 0.45 & 0.50 & 0.47 & 0.40 & 0.45 \\ \hline
\textbf{16} & 0.50 & 0.46 & 0.60 & 0.43 & 0.50 & 0.45 & 0.60 & 0.44 & 0.50 & 0.47 & 0.60 & 0.44 & 0.50 & 0.47 & 0.60 & 0.44 & 0.50 & 0.47 & 0.60 & 0.44 & 0.50 & 0.47 & 0.60 & 0.44 \\ \hline
\textbf{17} & 0.50 & 0.46 & 0.60 & 0.42 & 0.50 & 0.45 & 0.60 & 0.42 & 0.50 & 0.47 & 0.60 & 0.42 & 0.50 & 0.47 & 0.60 & 0.42 & 0.50 & 0.47 & 0.60 & 0.42 & 0.50 & 0.47 & 0.60 & 0.42 \\ \hline
\textbf{18} & 0.50 & 0.46 & 0.60 & 0.40 & 0.50 & 0.45 & 0.60 & 0.42 & 0.50 & 0.47 & 0.60 & 0.43 & 0.50 & 0.47 & 0.60 & 0.43 & 0.50 & 0.47 & 0.60 & 0.43 & 0.50 & 0.47 & 0.60 & 0.43 \\ \hline
\color{black}{\textbf{19}} & \color{black}{\textbf{0.50}} & \color{black}{\textbf{0.46}} & \color{black}{\textbf{0.60}} & \color{black}{\textbf{0.47}} & \color{black}{\textbf{0.50}} & \color{black}{\textbf{0.45}} & \color{black}{\textbf{0.55}} & \color{black}{\textbf{0.47}} & \color{black}{\textbf{0.50}} & \color{black}{\textbf{0.47}} & \color{black}{\textbf{0.52}} & \color{black}{\textbf{0.47}} & \color{black}{\textbf{0.50}} & \color{black}{\textbf{0.47}} & \color{black}{\textbf{0.50}} & \color{black}{\textbf{0.47}} & \color{black}{\textbf{0.50}} & \color{black}{\textbf{0.47}} & \color{black}{\textbf{0.47}} & \color{black}{\textbf{0.47}} & \color{black}{\textbf{0.60}} & \color{black}{\textbf{0.47}} & \color{black}{\textbf{0.47}} & \color{black}{\textbf{0.47}} \\ \hline
\textbf{20} & 0.50 & 0.46 & 0.60 & 0.42 & 0.50 & 0.45 & 0.60 & 0.42 & 0.50 & 0.47 & 0.60 & 0.42 & 0.50 & 0.47 & 0.60 & 0.42 & 0.50 & 0.47 & 0.60 & 0.42 & 0.50 & 0.47 & 0.60 & 0.42 \\ \hline
\end{tabular}}
\end{table*}

\vspace{0.6em}
%
% The next two lines define the bibliography style to be used, and the bibliography file.
\bibliographystyle{unsrtnat}
\bibliography{sample-sigchi}

\end{document}